\documentclass[superscriptaddress,twocolumn,showpacs,amsmath,amssymb,floats]{revtex4}

\usepackage{graphicx}

\usepackage{dcolumn}

\usepackage{bm}

\usepackage[usenames]{color}

\begin{document}

\newcommand{\brm}[1]{\bm{{\rm #1}}}
\newcommand{\tens}[1]{\underline{\underline{#1}}}
\newcommand{\mm}{\overset{\leftrightarrow}{m}}
\newcommand{\xv}{\bm{{\rm x}}}
\newcommand{\Rv}{\bm{{\rm R}}}
\newcommand{\uv}{\bm{{\rm u}}}
\newcommand{\nv}{\bm{{\rm n}}}
\newcommand{\Nv}{\bm{{\rm N}}}
\newcommand{\ev}{\bm{{\rm e}}}
\newcommand{\bv}{\bm{{\rm b}}}
\def\ten#1{\underline{\underline{{#1}}}}
\newcommand{\Ft}{{\tilde F}}
\newcommand{\Ftv}{\tilde{\mathbf{F}}}
\newcommand{\sigmat}{{\tilde \sigma}}
\newcommand{\sigmab}{{\overline \sigma}}
\newcommand{\ellv}{{\mathbf \ell}}
\newcommand{\qv}{\bm{{\rm q}}}
\newcommand{\pv}{\bm{{\rm p}}}
\newcommand{\tD}{\tilde{D}}

\title{Elasticity and Response in Nearly Isostatic Periodic Lattices}
\author{Anton Souslov, Andrea J. Liu and T. C. Lubensky}
\affiliation{Department of Physics and Astronomy, University of
Pennsylvania, Philadelphia, PA 19104, USA }

\date{\today}

\begin{abstract}
\noindent The square and kagome lattices with nearest neighbor
springs of spring constant $k$ are isostatic with a number of
zero-frequency modes that scale with their perimeter. We
analytically study the approach to this isostatic limit as
the spring constant $k'$ for next-nearest-neighbor bonds vanishes.
We identify a characteristic frequency $\omega^* \sim \sqrt{k'}$ and
length $l^* \sim \sqrt{k/k'}$ for both lattices. The shear modulus
$C_{44}= k'$ of the square lattice vanishes with $k'$, but that for
the kagome lattice does not.
\end{abstract}


\maketitle

An argument advanced by Maxwell~\cite{Maxwell1864} states that
rigid assemblies of loose particles must have at least $Z_{\rm
iso}=2n_f$ contacts per particle, where $n_f$ is the number of
relevant degrees of freedom per particle.   This result has been
applied to many systems \cite{Alexander1998,Wyart2005}, including
network glasses\cite{Phillips1981,Thorpe1983}, rigidity
percolation \cite{JacobsThor1995,DuxburyMou1999},
$\beta$-cristobalite \cite{SwainsonDov1993a}, granular media~
\cite{EdwardsGrin1999,TkachenkoWit1999}, protein
folding~\cite{RaderTho2002}, and elasticity in networks of
semi-flexible polymers \cite{HeussingerFre2007}.

Zero-temperature packings of particles can exhibit transitions
with increasing particle density from an unjammed, disordered
state with no inter-particle contacts to a jammed, disordered
state with an average of $Z_c$ contacts per particle.  These
transitions are discontinuous because the coordination number
$Z_c$ must satisfy Maxwell's inequality.  For the special case of
frictionless spheres, the coordination number at the jamming
transition, $Z_c$, appears to be exactly the minimal, or {\it
isostatic}, value needed for mechanical stability, $Z_{\rm
iso}$~\cite{Durian1995,OhernNag2002,OhernNag2003}.   The
coincidence of the jamming transition with the threshold for
mechanical stability gives rise to special properties at the
transition:  the coordination number $Z$ jumps discontinuously but
as the transition is approached from the high-density side, the
shear modulus vanishes and length and time scales diverge as power
laws~\cite{Durian1995,OhernNag2002,OhernNag2003,SilbertNag2005}.
In addition, the vibrational properties just above the transition
are very different from those of typical
solids~\cite{SilbertNag2005,WyartWit2005b,Wyart2005,XuNag2007,XuNag2009}.

\begin{figure}
\centerline{\includegraphics{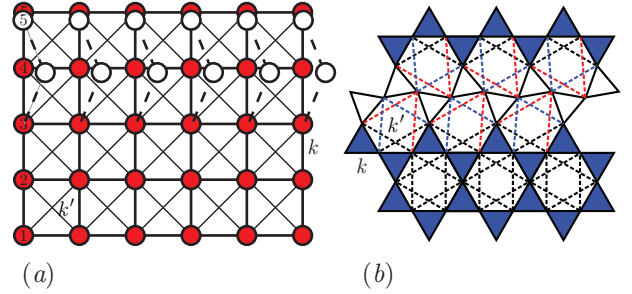}} \caption{(Color
Online) (a) Square and (b) kagome lattices with $NN$
springs of spring-constant $k$ and $NNN$ springs of spring
constant $k'$. White circles in (a) and white triangles in (b) show a zero-energy distortion.
} \label{fig:square-kagome1}
\end{figure}

The behavior of the isostatic jamming transition invites comparison
to second-order structural transitions, in which the vanishing of an
elastic constant in an isostatic lattice signals the instability
towards a shape distortion.  Jamming involves a transition between
two disordered states, while critical structural transitions involve
a crystalline state.  This raises the question of which aspects of
the jamming transition apply to {\it all} systems, periodic or
disordered, near isostatic transitions and which do not. To
answer these questions, it is important to study models for which
results can be rigourously established via analytic calculations.
In this paper, we test the robustness of the connection between
isostaticity, power-law elastic moduli and vibrational properties by
an exact analytical exploration of the approach to the
threshold of mechanical stability in fully periodic, nearly
isostatic systems. Specifically, we study two lattices, the $2D$
square and kagome lattices, shown in Fig.~\ref{fig:square-kagome1},
with nearest-neighbor ($NN$) harmonic springs of spring constant $k$
and next-nearest-neighbor ($NNN$) harmonic springs with
spring constant $k'$.  The isostatic structural transition is then
approached continuously as $k' \rightarrow 0$ since both systems are
isostatic there with $Z_c=4$. To describe the phase that
results when $k'<0$, it is necessary to add a nonlinear term $gx^4$
to the energy \cite{SouslovLub2009b}.

Our principal results are that in both the square and kagome
lattices, there is a characteristic frequency $\omega^* \sim
\sqrt{k'}$ and a length $l^* \sim \sqrt{k/k'}$ that
 follow directly form the form of the wavenumber- and
frequency-dependent response function but that can also be obtained
from cutting arguments \cite{SilbertNag2005,Wyart2005}. However, the
shear modulus exponent is different in the three
systems, showing that isostaticity does not confer universality on
all power-law properties near the transition.

It is useful first to understand the two states that straddle the
structural transition.  Above the transition, where $k'>0$, the
square/kagome lattice is stable.  Below the transition, where $k'<0$
(and $g>0$), the $N_0$ zero modes at $k'=0$ develop
positive or negative amplitudes, and there are $\sim a^{N_0}$, where
$a>1$, distinct ground states. In this paper, we will restrict our attention to the case $k' \ge 0$, and we will use the harmonic approximation.

Now consider the square lattice shown in
Fig.~\ref{fig:square-kagome1}(a) with $NNN$ nonlinear springs. This
lattice has $N=N_x N_y$ sites and $N_{nn} = 2 N_x N_y - N_x - N_y$
nearest-neighbor bonds.  Thus when $k'=0$, this system is isostatic
with a number of zero-frequency modes, $N_0 = 2N - N_{nn} = N_x +
N_y$, equal to half the perimeter of the system. When $k'>0$,
these modes become quasi-isostatic with nonzero frequencies
that vanish as $k' \rightarrow 0$. The components of the dynamical
matrix are easily calculated \cite{AshcroftMer}
\begin{align}
& D_{xx} ( \qv )  = D_{yy}(q_y,q_x)=  4 k \sin^2 (q_x /2)+ 4 k' \sin^2 (q_y /2) \nonumber\\
& + 4 k' \sin^2 (q_x /2) - 8 k' \sin^2 (q_x /2)\sin^2 (q_y /2) \nonumber  \\
& D_{xy} ( \qv)  = D_{yx}(\qv)= 2 k' \sin (q_x ) \sin (q_y ) ,
\label{eq:tensD-comp}
\end{align}
where we set the lattice constant $a$ equal to $1$.  In the
continuum limit $q \ll 1$,  $D_{ij} ( \qv)$ obtains the form
dictated by elasticity theory with elastic constants
$C_{11} = C_{xxxx} = C_{yyyy}$, $C_{12} = C_{xxyy}$ and $C_{44} =
C_{xyxy}$: \cite{AshcroftMer}
\begin{align}
& D_{xx}(\qv)=C_{11} q_x^2 + C_{44} q_y^2; \:\:
D_{yy}(\qv)=C_{11} q_y^2 + C_{44} q_x^2 \nonumber \\
& D_{xy}(\qv)=(C_{12}+C_{44}) q_x q_y, \label{eq:tildeC}
\end{align}
and we can relate $k$ and $k'$ to the elastic constants: $C_{11} = k
+ k'$, $C_{44} = C_{12} = k'$.  Thus, the shear modulus
vanishes as $k' \rightarrow 0$. When $k'=0$, $D_{ij} ( \qv)$ breaks
up into two independent one-dimensional compressional phonon
systems: it is diagonal with $D_{xx}(\qv) = 4 k \sin^2 (q_x/2)$
independent of $q_y$ and $D_{yy} ( \qv) = 4 k' \sin^2 (q_y/2)$
independent of $q_x$.  Thus at $q_x = 0$, $D_{xx}(\qv)$ vanishes for
all points along the line $-\pi < q_y < \pi$.  Note that at a
standard structural phase transition, components of $D_{ij} ( \qv)$
vanish at one or possibly a discrete set of points in the Brillouin
Zone (BZ) when an elastic modulus vanishes.  In contrast, for the
periodic isostatic system, components of $D_{ij}(\qv)$ vanish along
lines.

\begin{figure}
\centerline{\includegraphics{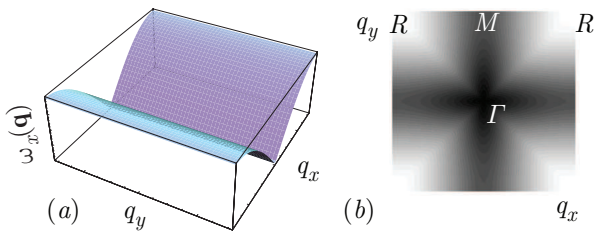}}
\caption{(a) $\omega_x(\qv)$ when $k'=0$ showing the line of zero
energy at $q_x=0$ and one-dimensional dispersion as a function of
$q_x$ at fixed $q_y$. (b) Density plot of the low-energy phonon
mode when $k' = 0.02k$ showing $\cos 4 \theta$ modulation at small
$\qv$ and one-dimensional isostatic behavior at large $\qv$.}
\label{fig:square-lattice2}
\end{figure}

\begin{figure}
\centerline{\includegraphics{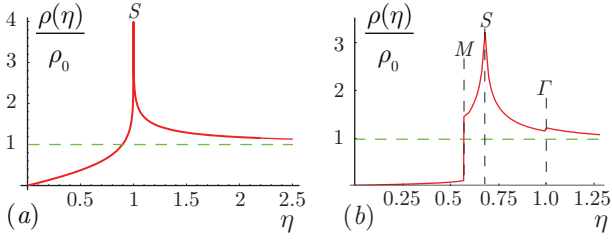}}
\caption{(Color Online) Density of states as a function of
$\eta=\omega/\omega^*$ for (a) the square and (b)
kagome lattices, showing their constant value when $k'=0$ (dashed
green line) and van Hove singularities, from the saddle at $S$
and minima at $M$ and $\Gamma$, whose frequencies all scale as
$\omega^* \propto \sqrt{k'}$ when $k' > 0$.}
\label{fig:sq-kagome-dos}
\end{figure}

The one-dimensional nature of $D_{ij} (\qv)$ gives rise to
compressional phonons with frequencies $\omega_{x,y} ( \qv ) = 2
\sqrt{k'} |\sin q_{x,y}/2|$ [Fig.~\ref{fig:square-lattice2}(a)]
and a one-dimensional density of states $\rho(\omega)
(2/\pi)/\sqrt{4 k - \omega^2}$ with a nonzero value $\rho_0=(\pi
\sqrt{k})^{-1}$ at $\omega = 0$ as shown in
Fig.~\ref{fig:sq-kagome-dos}(a).

When $k'>0$, modes exhibit a $\cos 4 \theta$ modulation at low
frequency and one-dimensional isostatic behavior at larger $\qv$
[Fig.~\ref{fig:square-lattice2}(b)]. When $0<k' \ll k$,
$D_{ij}(\qv)$ is well approximated as a diagonal matrix with
$D_{xx}(\qv) = k q_x^2 + 4 k' \sin^2(q_y/2)$ with associated
eigenfrequency $\omega_x(\qv) \sim \sqrt{D_{xx}(\qv)}$. These
expressions immediately define a characteristic frequency
\begin{equation}
\omega^*=2 \sqrt{k'} \label{eq:omsdefsq}
\end{equation}
as the frequency at the point $M=(0,\pi)$ on the BZ edge. The
first term in $D_{xx}(\qv)$ represents the long-wavelength
anomalous {\em isostatic (iso)} modes that are present when
$k'=0$, whereas the second represents the effects of $NNN$
coupling.  When $q_x=0$, the only length scale in the problem is
the unit lattice spacing, and no divergent length scale can be
extracted from $D_{xx}( 0 , q_y) $ as it can be in the case of the
structural phase transition. When the first term is large compared
to the second, $D_{xx}(\qv)$ reduces to its form for the isostatic
$k'=0$ limit, and we can extract a length by comparing these two
terms. The shortest length we can extract comes from comparing $k
q_x^2$ to $D_{xx}(\qv)$ at point $M$  on the zone edge, i.e.
\begin{equation}
l^* = (1/2)\sqrt{k/k'}\equiv 1/q^*. \label{eq:lstardefsq}
\end{equation}
If $q_y<\pi$, the isostatic limit is reached when $q_x > q^*$.  A
similar analysis applies to $D_{yy}(\qv)$ when $q_y > q^*$.  If a
square of length $l$ is cut from the bulk, the wavenumbers of its
excitations will be greater than $\pi/l$, and for $ql^*>1$, all
modes within the box will be effectively isostatic ones.  This
construction is equivalent to the cutting argument of Wyart et al.
\cite{WyartWit2005a,Wyart2005}.

Equation (\ref{eq:lstardefsq}) is identical to the length at which
the frequency of the compressional mode $\omega_x ( q_x, 0) =
\sqrt{k}/l^* ~\sim \sqrt{C_{11}}/l^*$ becomes equal to $\omega^*$.
A meaningful length from the transverse mode $\omega_x(0, q_y)$
cannot be extracted in a similar fashion.  The full phonon
spectrum [Fig.~\ref{fig:sq-kagome-dos}(a)] exhibits acoustic
phonons identical to those of a standard square lattice at $q\ll
1$ and a saddle point at the point $M$.  Thus, the low-frequency
density of states is Debye-like: $\rho(\omega) = (\omega/(2
\pi))/\sqrt{k k'}$ with a denominator that, because of the
anisotropy of the square lattice, is proportional to the geometric
mean of longitudinal and transverse sound velocities rather than
to a single velocity.  In addition $\rho(\omega)$ exhibits a
logarithmic van Hove singularity at $\omega^*$ and approaches the
one-dimensional limit $(1/\pi)/\sqrt{k}$ at $\omega^* \ll \omega
\ll 2\sqrt{k}$.  The frequency $\omega^*$
[Eq.~(\ref{eq:omsdefsq})] is recovered by equating the
low-frequency Debye form at $\omega^*$ to the high-frequency
isostatic form of the density of states.

The kagome lattice can be viewed as an array of one-dimensional
staggered linear rows, parallel to the $x$-axis, of pairs of
opposing triangles as shown in Fig.~\ref{fig:square-kagome1}.
Identical arrays with rows parallel to $(\cos 2 \pi/3, \pm\sin 2
\pi/3)$ can be identified. As in the square lattice, each site has
$2d=4$ nearest neighbors in the bulk. A counting procedure similar
to that used for the square lattice yields $N_0$ proportional to
the $\sqrt{N}$ for lattice of $N$ sites.

\begin{figure}
\centerline{\includegraphics{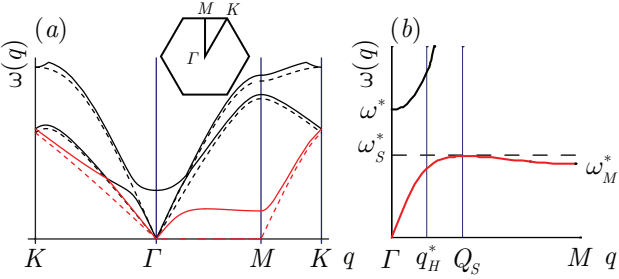}} \caption{(Color
Online) (a) Phonon dispersion along symmetry directions. The dotted
lines are for $k'=0$ and the solid lines are for $k'=0.02$. The
isostatic and quasi-isostatic branches are in red. (b)
shows isostatic and shear modes along $\Gamma M$ and indicates
characteristic frequencies and wavenumbers.}
\label{fig:kagome-modes-1}
\end{figure}

\begin{figure}
\centerline{\includegraphics{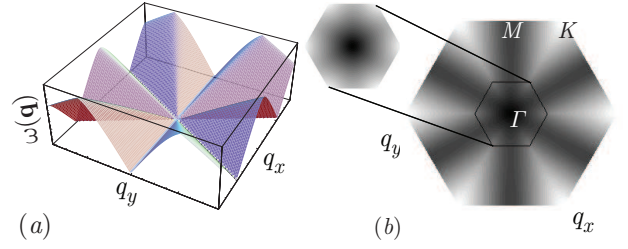}}
\caption{(a) Plot of iso mode at $k'=0$. (b) density plot of iso
mode at $k'=0.02 k$. Inset: circular symmetry at origin.}
\label{fig:kagome-modes-2}
\end{figure}

The isostatic (iso) modes correspond to identical rotations about
their top vertices of all of the ``up" pointing triangles in any
row in the horizontal (or symmetry-equivalent) grid.  These
rotations require counter rotations of connected ``down" pointing
triangles as shown in Fig.~\ref{fig:square-kagome1}(b). There are
three sites per unit cell (which we take to be those in the ``up"
triangles) in the kagome lattice and six phonon branches
[Fig.~\ref{fig:kagome-modes-1}(a)]. Three of these are high-energy
optical branches, two are acoustic and one is isostatic.  The zero
modes of the latter show up as three lines of zero frequency along
$q_x=0$ [$\Gamma =(0,0)$ to $M=(0,G_0/2)$, where $G_0 = 4
\pi/\sqrt{3}$, in the BZ] and the two symmetry-related lines as
shown in Fig.~\ref{fig:kagome-modes-2}(a). Away from $q_x=0$, the
isostatic mode frequency is $\omega_I ( \qv) = c_x q_x$, where
$c_x=\sqrt{3 k}/4$ for small $q_x$.  This behavior is identical to
that of the square lattice. The resulting density of states
decreases linearly with $\omega$ from a nonzero value $\rho_0 =
3G_0/(4 \pi^2 c_x)=8/(\pi \sqrt{k})$ at $\omega = 0$.  The total
low-frequency density of states from the two acoustic modes and
the isostatic mode is independent of $\omega$ and equal to
$\rho_0$ at small $\omega$ [Fig.~\ref{fig:sq-kagome-dos}(b)].

When springs of spring constant $k'$ are added to the $NNN$ bonds
shown in Fig.~\ref{fig:square-kagome1}(b), the
quasi-isostatic mode along $q_x=0$
[Fig.~\ref{fig:kagome-modes-1}(b)] has nonzero frequency of order
$\sqrt{k'}$ for all $q_y>0$ and gives rise to various lengths of
order $l^* \sim \sqrt{k/k'}$.  At $q_x=0$ and low values of
$q_y$, this mode hybridizes with the transverse phonon mode to
produce a gapped translation-rotation mode with frequency
\begin{equation}
\omega^* = \sqrt{6k'} \label{eq:omsdefkg}
\end{equation}
at $\qv = 0$.  At small $\qv$, there are isotropic longitudinal
and transverse sound modes with respective velocities $c_L=
\sqrt{3 k}/4$ and $c_T = \sqrt{k}/4$. The lowest frequency mode
after hybridization at $q_x=0$ is a
transverse phonon near $q_y = 0$ and predominantly an isostatic
rotation mode at $q_y>q_H^*$, where $q_H^* = 4\sqrt{3 k'/k} \equiv
1/l_H^* $ can be termed a hybridization wave number.  This mode
reaches a maximum frequency $\omega_S^* = \omega^*/\sqrt{2}$ at a
saddle point at $q_y = Q_S = 4 (3 k'/2 k)^{1/4} \sim (l^*)^{-1/2}$
and a local minimum with frequency $\omega_M^*=\omega^*/\sqrt{3}$ at
the zone-edge point $M$. At low frequency, the DOS is Debye-like:
$\rho(\omega)=(\omega/2 \pi)(c_L^{-2} + c_T^{-2}) = 32 \omega/( 2
\pi k)$. The points $M$, $\Gamma $ and $S= (0,Q_S)$ give rise to
van Hove singularities in the DOS.  The minimum point $M$ produces
a jump $\Delta \rho_M= 8 \sqrt{2}/\pi \sqrt{k}>\rho_0$ at
$\omega_M^*$, the saddle $S$ produces a logarithmic singularity at
$\omega_S^*$, and the minimum at $\Gamma$ a jump $\Delta
\rho_{\Gamma} = (16 /\pi) \sqrt{3 k'/2k} \ll \rho_0$, which is
just visible in Fig.~\ref{fig:sq-kagome-dos}(b).

Lengths that scale as $\sqrt{k/k'}$ can be introduced in much the
same way as in the square lattice.  The square of the
low-frequency isostatic-shear mode at $q_x=0$ increases as $c_x^2
q_x^2$ for nonzero shear, and lengths $l_S^* = (1/4) \sqrt{k/k'}$
and $l_M^* = (\sqrt{6}/8) \sqrt{k/k'} $ follow from comparing $c_x
q_x$ to $\omega_S^*$ and $\omega_M^*$, respectively. These two
lengths are longer than the hybridization length, $l_H^*$. Thus,
it is only at length scales less than $l_H^*$ that isostatic modes
are retrieved completely, and we should take
\begin{equation}
l^* = l_H^*= (1/4) \sqrt{k/(3k')}. \label{eq:lstardefkg}
\end{equation}
The less divergent length $\xi_S = Q_S^{-1} \sim \sqrt{l^*}$
determines the position of the saddle $S$ and does not describe
the same physics as the other lengths.

The long-wavelength, low-frequency properties of the
kagome lattice are best understood by considering the effective
low-energy dynamical matrix $\tD_{ij}$ obtained by integrating out
the high-energy modes. This matrix is conveniently represented in a
basis consisting of the longitudinal ($L$) and transverse ($T$)
phonons and the mixed rotation-translation gapped mode ($\eta$) at
$\qv $: $\tD_{LT}= \tD_{TL}=0$, and components are
\begin{eqnarray}
\tD_{LL}& = & 3 k q^2/16; \:\:\tD_{TT} = k q^2/16; \:\:
\tD_{\eta \eta} = 6 k' + k q^2/16 \nonumber \\
\tD_{L\eta}& = & \tD_{\eta L}^*= k q^2 \cos(3 \theta )/16 \\
\tD_{T\eta}& = & \tD_{\eta T}^* = - i (\sqrt{3}/2) k' q + k q^2
\sin (3 \theta)/16 , \nonumber
\end{eqnarray}
where $\theta$ is the angle that $\qv$ makes with the $x$-axis.
When $k'=0$, the transverse phonon and rotation-translation mode
mix at $q_x=0$ ($\theta= \pi/2$) to produce the zero isostatic
rotation mode and a sound mode with $\omega = \sqrt{k/8}q_y$. In
addition, there is the longitudinal mode with $\omega = (\sqrt{3
k}/4) q_y$. When $k'\neq 0$, the off-diagonal terms can be ignored
to lowest order in $\qv$, leading to longitudinal and transverse
sound modes with respective velocities $c_L = \sqrt{B}$ and $c_T=
\sqrt{G}$, arising from bulk and shear moduli $B= 3k/16$ and
$G=k/16$. Note that the shear modulus is proportional to $k$; this
is in contrast to the square lattice where  $G \propto k'$. To
understand how this comes about, it is instructive to look at the
$TT$ component of the phonon susceptibility,
\begin{equation}
\chi_{TT}(q, \theta) = D_{TT}^{-1}(q,\theta) = \frac{16}{k
q^2}\frac{ 6 + (q l^*)^2 (3-\cos^2 3 \theta)}{6+ 2 (q l^*)^2
\cos^2 3 \theta } . \label{eq:chiTT}
\end{equation}
When $ql^* \ll 1$, this reduces to the required isotropic form
$1/(G q^2)$.  When $ql^* \gg 1$, $\chi_{TT}$ is anisotropic with
value $1/(Gq^2) $ for $\cos 3 \theta = 1$ and $1/(6 k')$ for $\cos
3 \theta = 0$.

\begin{table}
\caption{\label{table} Frequencies, lengths, and
elastic moduli in square and kagome lattices and the
marginally-jammed (MJ) state of disordered sphere packings.  $\omega^*$, $l^*$ and $B$ scale the same way in all lattices
if we take $k' \sim (\Delta z)^2$, but $G$ does not.}
\begin{ruledtabular}
\begin{tabular}{|c|ccc|}
\hline Quantity & Square & Kagome & MJ
\\
\hline $l^*$ & $\left(\frac{k}{k'}\right)^{1/2}$ &
$\left(\frac{k}{k'}\right)^{1/2}$ & $\left(\Delta z\right)^{-1}$
\\
$\omega^*$ & $\left(k'\right)^{1/2}$ & $\left(k'\right)^{1/2}$ &
$\left(\Delta z\right)^1$
\\
$G$ & $k'$ & $k$ & $\left(\Delta z\right)^1$
\\
$B$ & $k$ & $k$ & $\left(\Delta z\right)^0$
\end{tabular}
\end{ruledtabular}
\end{table}

We can now compare the properties of nearly isostatic lattices
with those of disordered sphere packings.  Marginally-jammed
packings with volume fraction $\phi \equiv \phi_c + \Delta \phi$,
just above the volume fraction $\phi_c$ at the jamming threshold,
have an average number of contacts per sphere $z\equiv z_c +
\Delta z$, where $z_c = 2 d$, and $\Delta z \sim(\Delta
\phi)^{1/2}$~\cite{Durian1995,OhernNag2003}.  They are
macroscopically isotropic with bulk and shear moduli that for
harmonic inter-particle potentials scale respectively as $B\sim
(\Delta z)^0$ and $G\sim (\Delta
z)^1$~\cite{Durian1995,OhernNag2003}. The longitudinal and
transverse sound velocities then scale as $c_L \sim (\Delta z)^0$
and $c_T \sim (\Delta z)^{1/2}$.  The density of states has a
plateau above a frequency
$\omega^*$~\cite{OhernNag2003,SilbertNag2005}.  A length scale
$l^*\sim 1/\Delta z$ can be extracted by equating $c_L/l^*$ to
$\omega^*$~\cite{SilbertNag2005}, or from the response to a point
perturbation~\cite{EllenbroekSaa2006}. This scaling emerges from
the cutting arguments of
Refs.~\onlinecite{WyartWit2005b,Wyart2005}. A second length
$l_T^*\sim(\Delta z)^{-1/2}$ can be defined from $c_T/l_T^* =
\omega^*$~\cite{SilbertNag2005} and is important for energy
transport~\cite{XuNag2009}.

The most robust features are the scalings of
$\omega^* \sim \sqrt{k'}$ and $l^*\sim 1/\sqrt{k'}$ in both the
square and kagome lattices.  Effective medium calculations
\cite{MaoLub2009} for these lattices with $NNN$ bonds added with
probability $P \sim \Delta z$ yield $k'\sim(\Delta z)^2$; this
correspondence yields the observed scalings of $\omega^*$
and $l^*$ for marginally-jammed systems.  These
scalings are likely to be robust for all nearly-isostatic systems.
However, the power-law scaling of the shear modulus
is not universal (Table \ref{table}).  The
shear moduli of marginally-jammed systems and the square lattice
both vanish with $\Delta z$ (with different powers), but $G$ of
the kagome lattice is proportional to $k$ and does not vanish.
Thus, different isostatic transitions can fall into
different universality classes. This conclusion is consistent with
recent numerical results for disordered systems near isostaticity,
which show different scalings of bulk moduli~\cite{EllenbroekSaa2009}.

We thank NSF-DMR-0520020 (AS), NSF-DMR-0804900
(TCL), and DE-FG02-05ER46199 (AJL).

\end{document}